# DEFENDING AGAINST CYBERSECURITY THREATS TO THE PAYMENTS AND BANKING SYSTEM

Williams Haruna
*Cybersecurity and Human Factor*
Bournemouth University
Bournemouth, United Kingdom
s5433111@bournemouth.ac.uk

Toyin Ajiboro Aremu
*Cybersecurity and Human Factor*
Bournemouth University
Bournemouth,United Kingdom
s5527763@bournemouth.ac.uk

Ajao Yetunde Modupe
*Cybersecurity and Human Factor*
Bournemouth university
Bournemouth, United Kingdom
s5537712@bournmouth.ac.uk

*Abstract*— Cyber security threats to the payment and banking system have become a worldwide menace. The phenomenon has forced financial institutions to take risk as part of their business model. Hence, deliberate investment in sophisticated technologies and security measures have become imperative to safeguard against heavy financial losses and information breaches that may occur due to cyber-attacks. The proliferation of cyber-crimes is a huge concern for various stakeholders in the banking sector. Usually, cyber-attacks are carried out via software systems running on a computing system in cyberspace. As such, to prevent risks of cyber-attacks on the software systems, entities operating within the cyberspace must be identified and the threats to the application security isolated after analyzing the vulnerabilities and developing defense mechanisms. This paper will examine various approaches that identify assets in cyberspace, classify the cyber threats, provide security defenses and map security measures to control types and functionalities. Thus, adopting the right application to the security threats and defenses will aid IT professionals and users alike in making decisions for developing a strong defense-in-depth mechanism.

*Keywords—Threat, Risk, cyber-attack, financial institution, (key words)*

## I. INTRODUCTION

Around the world, banking and payment applications have made it simpler for regular people to conduct transactions on their smartphones and devices from the comfort of their homes. Due to the streamlined methods of completing financial transactions, the banking industry has rapidly developed over the past few decades, resulting in the unification of information and computer networks into one information and cyber space. The ease with which banking transactions may be carried out accounts for the phenomenon's adoption on a global scale. Users of mobile banks can, for instance, carry out banking-related tasks including balance queries, transfers, and payments without going into a bank. Users appreciate getting alerts from the bank about recent large transactions, low balances, and overdrafts, among other things.

Since banking and payment are frequently conducted virtually, it is now the end user's responsibility to do their own security checks and the service provider to create a secure platform. Banking and payments can be categorized into three types, namely browser, application, and messaging [1]. The screen lock is therefore the first line of defense while banking on a smartphone. When physical access is used to get around that, the screen locks password is no longer secure. It is safe to say that the lock screen, which serves as the initial line of protection, offers minimal security.

## II. CYBER SECURITY AND THE FINANCIAL INSTITUTION

Asia-based businesses accounted for more than 81 billion of the 450 billion dollars in costs associated with cybercrime in the global economy. High profile cyberattacks frequently include denial of service, infrastructure attack, phishing, and other data protection-related problems [2]. The capital market and banks have experienced several CEO whaling assaults, which are viewed as a threat to the industry's cyber security. Financial service businesses have been affected by incidents focusing on cyber security far more than organizations in other sectors have. On the other side, 33% of major attacks target the financial services sector. In light of this, it is crucial to create a few security procedures to guard against cyberthreats in the banking industry. Ransomware attacks have recently had a significant impact on financial institutions. Artificial intelligence and machine learning are being used to thwart hackers.

### A. Security Threat and Banking and Payment

With the evolution of banking from traditional paper and pen to making transactions to the use of computers, mobile devices, and other gadgets has opened up various channels and threats to the system, making it unsafe for both the institution and individuals, this paper sheds light on the growing problem of cyber security as it relates to payment and banking systems. There are five main categories of security concerns that can harm the banking and payment system, according to Alzoubi et al. in 2022. The banking and payment systems are susceptible to five different kinds of security problems.

Although the enhanced development of commuting technology offers numerous benefits, they also present new difficulties. Common problems like fraud and theft are being committed in new kinds of cybercrime through information technology [3]. The types of cybercrimes are expanding constantly, and information technology helps and facilitates them regardless of continent or country. As a result, it is now a transnational crime. Monitoring, detecting, preventing, and controlling are made more challenging by this. For instance, ransomware, denial of service attacks, and phishing all directly affect commercial networks. Because it is a behavioral pattern for various accounts, this is challenging to identify. These includes

1. Unencrypted data: The confidence a client has in the system is that their data, including pin codes and credit card numbers, is safe and secure. This keeps happening because most people have no knowledge how to secure their data. Most of the time, the data are not safeguarded, making it simple for malicious attackers to use them to steal money from the victim's account. [4]

2. Malware: This popular hacking technique involves sending harmful malware into users' devices over the internet, usually by hackers.[5], hacked computers and cell phones constitute a serious threat to the banking and payment systems. The drawback of utilizing such software is the ease with which hackers can get access to the system and steal substantial sums of money from banks and payment gateways without leaving a trace or being discovered.

3. Unreliable third-party services: The banks use third-party suppliers who offer superior services to the payment and banking systems. However, if the third party's system is open to intrusion, it will be simple for it to be infiltrated, which could result in theft using the compromised third-party system. [6], the main outcome of this will be the destruction of the bank's reputation.

4. Manipulated data: When data related to cyber security systems are altered by hackers, it makes it easier for them to trick individuals into handing them money under false pretenses. [6]. claimed that the bank and payment system will suffer financial losses as a result.

5. Spoofing: This is more like pretending to be someone you are not, or impersonation. Hackers use this method to pretend to be the account's owner. To accomplish this, they obtain a person's login information and use an unlawful login to steal from the victim.{7] this is more detrimental to the individual than the banks.

These points have provided details on the issue of security in banks and other financial institutions. Additionally, it demonstrates how the banking industry's flaws and methods of operation effect consumers and how nefarious individuals exploit them.

This provides a general overview of why hardening and implementing cyber security measures should be taken into consideration, especially by those that interface with such systems.

**Cases of Cyber-Attacks on Financial institutions Globally**

The New York Times reports that a cyber-attack in 2012 resulted in angry customers of Bank of America, Chase Bank, Citigroup U.S. Bank, Wells Fargo, and PNC who were unable to access their accounts. The bank was unable to explain in detail the circumstances or what was occurring [8]. However, the CEO then revealed to analysts that the bank spent millions of dollars annually on cyber security to guard against data breaches [9] and that this was a case of denial of service that was intended to result in financial loss for the institution. The attacker wanted to temporarily disable the bank's public-facing website [10].

According to a 2014 article in USA Today, federal officials warned businesses on Monday that hackers had stolen more than 500 million financial records over the course of the previous year by attacking banks [11]. In 2016, it was also claimed in another news source that DDOS assaults against 46 large financial institutions had been launched. In these operations, hackers had gained remote access to hundreds of computers and servers, flooded them with data, and blocked lawful traffic. The attack struck JP Morgan the most severely, and more than 83 million of the bank's customers were compromised [11].

Europe has also experienced its fair share, as demonstrated by the cyberattack on the RBS banking group's online payment system in 2015 that prevented customers from logging in as their paychecks were entering their accounts [12]. As reported by NASDAQ, various cyberattacks occurred on online trading platforms throughout 2015. One such incident involved the data breach of FXCM, an online foreign exchange trading platform where unlawful transfers were made [13].

A trojan known as corkow (Metel) took control of the stock trading terminal and placed orders totaling several hundred million dollars, according to a report issued in 2015 by the IB Group, an information security organization. Dollar was bought for 55 rubles and sold for 65 rubles, causing extraordinary volatility. Russian bank was impacted, suffering significant losses Additionally in 2015, hackers attempted to use the SWIFT system to steal $951 from the Bangladeshi state bank. Also in 2016, hackers from the Lurk team, which develops the Trojan, were caught by the interior ministry after successfully stealing 1.7 billion rubles, or 28.3 million USD, from Russian banks.

Another cyberattacks, called Buhtrap, saw thefts from Russian banks range in value from $370,000 to $9 million. Additionally, HSBC, one of the biggest banks in the world, experienced a cyber-attack that prevented consumers from using online banking services twice a month [14].

A phishing attempt on the ICICI bank in Asia resulted in a well-known court case for fraud brought by a client. Additionally, the three largest banks, ICICI Bank, HDFC Bank, and others, reported that some of their client card accounts may have been compromised following the use of an ATM [15]. In Russia, the hacker organization Lazarus is well known for stealing $81 million from the Bangladesh Bank in 2016.

In Turkey, an insurance magazine reported that hackers specifically targeted AKBank in a cyberattack on the Swift system for international money transfers. According to the bank, the incident might result in liability of up to $4 million. According to a 2017 Taipei Times article, hackers moved $60 million from the Far Eastern Bank to accounts in Sri Lanka, Cambodia, and the United States as a result of malware that had been implanted in the system of the bank. This virus affected PCs, servers, and the Swift network [16]. 2018 saw Habib Bank Limited suffer from ATM skimming, which resulted in the theft of approximately Rs10 million from 559 of their accounts.

At least 19 firms in Kenya were reportedly impacted by the ransomware virus in a global attack in 2016, which also targeted 10 institutions in the banking, insurance, and government sectors across three countries in Africa. Nearly $206 million was lost as a result of this attack. [17]

Additionally, Australia's central bank was a target of a cyberattack in 2016. A bogus common wealth scam has targeted thousands of Australians. Bank emails that include malware are sent to recipients. The hoax, which urges visitors to click to see a "Secure Message," targets both customers and noncustomers [18]. Additionally, people who fall for the trap will actually download a trojan that hackers employ to break into systems [19].

## III. IDENTIFYING RISK AND THE THREAT LEADING CYBERCRIMES

Before An event or activity that could have a negative consequence or outcome is referred to as being at risk. Taking a risk could have a positive or negative outcome. Risk is merely a term for a component of uncertainty or the possibility of a variety of potential outcomes. [20]. Risks to banking and payment systems include ransomware, denial of service, race condition, phishing, and data breach.

### A. Ransomwares

Use One of the cyber dangers with the greatest growth targets a variety of users, including home users, students, and corporate networks. Newsweek reports that since January 2016, there have been an average of 4000 ransomware assaults each day, costing billions of dollars.

Data from a credit card holder will be held hostage by ransomware until the ransom is paid and the control is restored.

It enables attackers to quickly earn a reward—typically a modest money, but occasionally a substantial number—and then move on to the target victim. Attackers get access to a system by leveraging email attachments, hyperlinks included in email messages, and application or internet security flaws. The malware that encrypts files, discs, and locks down systems is installed when an employee of the institution clicks on harmful files. The attacker sends a message requesting payment to decrypt the data, and because they now accept payment in crypto currencies, it is challenging to trace them.

### B. Phishing and business email compromise

More than $5 billion had been stolen from businesses through email hacks, including phishing, according to a May 2017 FBI investigation. When a criminal pretends to be a reliable source in order to obtain information or money, this is known as phishing. There are many types of phishing, including spear phishing and whaling. The attacker poses as the CEO or a representative of the business and demands money or sensitive data. If banks and payment systems do not identify this attacker, they run the danger of suffering financial losses and compromising client information, of the recorded 37.3 million

### C. Watering hole attacks

The Watering hole attacks have grown in popularity as a means of compromise as employee understanding of phishing emails has improved. An attacker utilizes a compromised website that they know or suspect their victim will visit in a "watering hole attack" to infect their system with malware. Websites relating to the financial sector are increasingly being compromised. The website of the Polish Financial Supervision Authority was found to have been infiltrated in February 2017 by a significant watering hole assault. The investigation's first findings indicate that the intrusions were likely exploited to steal data from the banks' networks. The attack targeted more than 100 financial institutions in 30 countries. 60 The RGB/Lazarus Group has been implicated in the assault. The training that bank employees receive allows attackers to infect targets who have been educated not to click links, grant remote access to their workstations, or even open emails from strangers. One bank executive also highlighted a new type of watering hole assault that is emerging. Instead, many businesses advise staff to Google the identities of individuals or businesses who cold email them to make sure they are real before communicating with them. Therefore, some attackers send phishing emails to their targets while also developing compromised websites for the company that is purportedly the sender of the email in the hopes that the employee will visit the website to confirm the email's legitimacy.

Attacks are evolving Two new attack types that emerged in 2016 are a major concern for bank cybersecurity teams. All bank executives, law enforcement officers, and experts consulted for this analysis recognized DDoS attacks from internet of things (IoT) botnets as the top emerging danger. The majority also concurred that ransomware assaults were the second-biggest threat.

## IV. HUMAN FACTOR ASPECT OF CYBERSECURITY ON PAYMENT AND BANKING

Humans are using the internet more frequently, which increases our vulnerability to attacks and has an effect on the devices we use to access online banking and make payments, which are typically set to the lowest level of security, requiring users to do it safely. Terms of agreements, intangible data, and increasingly complicated technologies all make it challenging to maintain the privacy of our identity. The negative outcome is the loss of sensitive data, including login information, which can be harmful to us. As humans use connected technology and infrastructure to interact and share data, they have varied levels of security.

## V. LEGISLATION ETHICAL STANDARD AND CYBER SECURITY

Laws are rules that provide specific parameters within which stakeholders may do their business. Cybercrime is typically committed using a variety of devices via open networks in an effort to obtain certain rewards. Information continues to be the most valuable resource on the internet. As a result, the law exists to specify what is legal and what is illegal when using the internet. Additionally, it outlines the offences and penalties associated with specific online infarctions. The laws and regulations needed to safeguard the information system and assets under a certain authority, however, are drafted by the legislative. The legislature alone is responsible for developing cyber law legislation and policy. Regarding financial institutions, breaking any legislation has serious consequences in terms of fines and penalties, and those who commit these crimes will be penalized after being successfully prosecuted by the criminal justice system. The following laws have been implemented by various nations throughout the world to safeguard banking systems:

Unilateral countermeasures for peacetime: This law is designed to thwart criminals who would target and harm information assets through cybercrime. Countermeasures are ways for parties to international law to impose the forceful enforcement of their rights. Therefore, an employee's rights under the banking system serve as a guarantee that they won't transgress company policy and guidelines.

Sanctions by the Security Council

Because they target both the assets and clients of the bank, cyberattacks pose a serious danger to the continued existence of online banking and payment systems. As stated in Article 5 of the International Cyber Security Law, incidences of cyberattacks do not qualify as armed attacks. Its main purpose is to persuade the Security Council to detain and charge the aforementioned offender. However, the Security Council may impose penalties on the offending nations or people if there is a direct cyberattack against it.

Cyber Law Enforcement Cooperation: This guideline outlines the types of online behavior that are acceptable for people to engage in. As a result, every person and international organization must adhere to all prescribed conducts as outlined in the treaty.

Proper cyber hygiene

This refers to the procedure of keeping an eye on and analyzing the actions that result in security breaches so that appropriate action can be taken. This process is ongoing. This approach, which is synonymous with asset protection for banks and operational resilience, is not a law in and of itself. These are now considered ethical dilemmas in the corporate and banking sectors all over the world.

The Challenge of Global Cooperation

Information system cybersecurity breaches offer hazards with a global scope. The majority of banks are multinational, which means they all have online presences and have locations all over the world. Since the internet is intended for a specific location, attackers are also not headquartered in a single location or nation. Because dangers are global in scope, bankers must adopt an approach when taking action. Since attackers from nations with less stringent cybersecurity measures still pose a threat to banks in other climes, it is not enough for a specific country to have efficient cyber security laws to discourage criminals in their territories. As a result, all security measures must be global in scope. Although a global international legislation governing and guiding cyber security would have been ideal, in practice only proactive steps using technology instruments are required to secure information assets and applications.

Ethical Issues

Employees of banks are said to be affected by ethical difficulties. Therefore, it is regarded unethical when an employee goes above his level of security access to get unlawful information. Employees had to get higher-ups' approval before revealing any material due to ethical considerations. This transcends domain function and has an impact on personal functionalities. Lesion domain issues are not the only ethical ones. (Banking-related moral concerns. While ethical concerns are related to a person's behaviour and day-to-day employment. The banking and payment systems raise four (4) ethical concerns. These are what they are:

Privacy : Privacy and confidentiality are the same thing. For the benefit of their owners, all general and sensitive information must be kept private. Any revelation made without the owners' express consent is unethical. Information that is confidential or extremely sensitive cannot be shared with anybody but the owner. Customers of the banking and payment systems must possess specific credentials in order to use their services. The credentials are private information that bank workers are not permitted to share. The existence of the entire banking sector and the confidence of their consumers are gravely threatened by the actions of a few dishonest personnel and internet attackers.

Accuracy: Accurate information is essential since false information could mislead a client or any bank or payment system workers. This will make it more difficult for the information to be provided to transactional and benchmarking systems for functional feasibility. As a result, the system cannot function due to the inability to interpret any reliable information, which lowers public confidence in it.

Property: Intellectual property is a significant problem for both societal organizations and financial enterprises. Laws like the Copyright Act, patent oaths, and confidentiality oaths are an outdated system for defending intellectual property rights. In order to safeguard users of the digital era from attackers and disturbances frequently encountered online while using bank applications, intellectual property must be protected. Banks presently use digital signatures to regulate ethical behaviour through electronic document signing.

Information access: Timely access to information is a moral concern because it would be unethical for someone to have access to someone else's information. There has to be an access method to prevent this from happening. Additionally, both customers and employees need to be literate in order to do their jobs. In order to prevent unwanted access to another person's information, the system must be reliable enough.

VI. INFORMATION AND PHYSICAL SECURITY OF BANKING AND PAYMENT SYSTEMS.

Information security is the most important component of the online banking payment system. By safeguarding the data entering and leaving the financial system that would be kept in the bank's database, it ensured secrecy. The requirement for mitigation arises from the risk of unauthorized access, usage, disclosure, disruption, and change of sensitive data in the financial system. Therefore, the deployment of the following technological instruments for the organization's robustness against attackers becomes essential to enable the development of a secure environment within the banking network..

.A. Processes and Security Services; under the functional domain of the bank, the following processes are carried out to ensure the security of private and sensitive information belonging to the bank and its clients:

I.Authentication: By using this procedure, all problems with accessing the bank's information and all intended functionalities are fixed. This specified users who could access the bank's computer network. It is a method used in conjunction with a secure password. One Time Password is used as the second layer of protection after that (OTP). By requiring a password that is unicasted to the user's mobile number on file with the bank, OTP makes the authentication procedures more onerous. The platform will only allow confirmed users access.

II. Authorization: This establishes the portion of the information that the verified user may access. This is the implementation of the access privileges and rights granted to bank workers and clients within the open network interface. Customers typically only have access to their bank accounts, and workers are only allowed to carry out tasks that have been approved by the bank. Typically, the bank's information technology officers work with their management to plan and implement this

III. Auditing: By law, all financial institutions must perform audits. Therefore, it is expected of privileged staff to conduct time-based audits of all operations conducted on the bank's servers and database. Audit trails are typically compared to the backed-up devices, which store the value of the bank both before and after a transaction. Here, details like the transaction date, terminal identity, user identification, name, bank's domain name, time stamp, and whether the transaction was successful or unsuccessful are logged.

IV. Confidentiality: The banking and payment system treats all information it processes as confidential. Therefore, effective benchmarking is necessary to value the knowledge of the bank's staff and clients. To do this, banks typically use several layers of encryption techniques to safeguard critical data. This applies to both symmetric or secret key cryptography and asymmetric or public key cryptography and is known as IPSec and Transport Layer Security (TLS)

V. The most important component of the banking and payment system is integrity. Simply said, it means that all user data must be same at the receiver's end. Data transmission over the internet can be intercepted by an outsider in an unsecured system by altering the data before it reaches the recipient. Similarly, the data kept in the database needs to be secured to stop attacks from altering it in any way. Data integrity is typically safeguarded using a digital signature and the Message Authentication Code (MAC). Information from the source is guaranteed to be accurate by maintaining data integrity. The banking system's success depends on this.

VI. Availability refers to the requirement that banking services be accessible at all times. An attacker might inject unnecessary packets into the banking system network to clog it up. Employees and customers of the bank may not be able to access the servers when this kind of issue occurs. Therefore, a reliable system is required to safeguard the bank against DOS and DDOS attacks. To ensure that banking services are constantly accessible, the majority of financial institutions often use techniques like intrusion detection systems (IDS), firewalls, and demilitarized zones (DMZ).

B. An incredibly powerful tool for network monitoring is PRTG Network Monitoring. It is popular with banking system stakeholders and simple to configure on the system. It offers a user interface that warns people when an event is about to happen. Users have a high level of trust in it and use it with administrator rights. The programme is able to find harmful activity on certain networks. This tool is used by bankers to oversee the security of various components of the banking and payment systems and applications.

## VII. HOW TO MITIGATE CYBER SECURITY THREATS IN BANKING AND PAYMENT SYSTEMS

Training and awareness must be the primary solutions for the issue of cyber security in digital banking. It is impossible to overstate the importance of educating employees and customers, as what you don't know poses a hazard that cannot be eliminated.

The importance of authenticity was also stated by Alhosani, F. A., & Tariq, M. U. in 2020[21]. That system will feature a number of authentication mechanisms that may be scanned to determine a person's identification and include a variety of data.

Banks and payment systems should implement

OTP: one-time password tokens to allow users access online services with the security level of OTP. According to Bose, R., Chakraborty, S., and Roy (2019)[22], the OTPs won't be in effect for very long.

Device identification is important since it helps to protect against hackers and attackers. The device that a person uses should also be authenticated. from using their information improperly.

monitoring of transactions and limit: [22] banks are presently using this security system, which also

## VIII. RELATED WORK

Numerous earlier studies have been written about the problem of cyber security in banking and payment systems. One of the works about difficulties with data security when using cloud computing in digital banking was done by Giri & Shakya [4]. Despite not focusing on a specific case study, the particularly focuses on such concerns and their occurrence in Nepal. The writers' secondary research and data-gathering techniques. This is extremely similar to the research approach used in this paper, which likewise employs secondary sources to gather data from earlier works of literature. Contrary to that, this paper just discusses the data rather than attempting to formulate any assumptions.

Hua et al[23] .'s research focuses on how resilient each person is to cyber security risks in digital banking.

This subject is also covered in a paper by Li et al. [5] regarding the cyber security measures that banks provide to ensure client happiness. Although this paper includes more instances than the article, Tariq's [21] article on cyber security concerns also uses secondary methodologies and quantitative analysis.

Additionally, Alzoubi et al[24]. examined the threat to digital banking using theories and earlier research done by other authors. In addition, Mary [25]. examined the human element in regard to banking systems' cyber security. Pradeep evaluated the legal and ethical requirements for information security.

## IX. CONCLUSION.

This study focuses on the elements that affect financial institutions' knowledge of threats and the solutions to these problems. As a result, the research's findings can be applied to other banks and payment systems, both nationally and internationally, as well as to any organization, and can act as protection against cyber risk and other types of attacks in general.

The rise to online banking and services has made cyber security even more crucial since the COVID-19 pandemic. According to the analysis, the majority of the large banks have both a cyber security policy and cyber insurances in place. Despite the fact that bank cyberattacks are becoming more extensive than ever. It remains to be seen how effective the tactics used by banks will be. Despite the significance of banking and online systems, there are currently few security studies in this field. This is despite the importance of these systems' awareness of threats, laws, and human factors as well as their readiness to defend themselves against attacks by adhering to industry standards and mitigation procedures

Cybercrime knows no boundaries and is developing at the same rate as technology. Since banks are one of the main economic cornerstones of every nation, it is essential to defend them against cyberattacks. The majority of banks and payment institutions must abide by the legislated framework and rules for cyber security and resilience. In recent years, there has been a shift in the banks' attitudes toward cybercrime. Banks are actively defending their assets against fraudsters and hackers, as well as the data of their customers.

This paper proposes that all Banking and Payment systems need to be better protected against cyberattacks, which is in line with the current cyber threat landscape. This is as a result of attackers' current emphasis on monetary rewards over bragging rights. Therefore, banking and payment institutions must be aware that cyberattack perpetrators may adopt cyber security measures for intelligence gathering on sensitive customer information and performing online fraudulent financial transactions in order to offer secure online financial services and maintain their good reputation. In the absence of such measures, financial services risk disruption, local financial system deterioration, and eventual economic collapse of the nation.